# Semiconductor Quantum Computation

Xin Zhang,[1,2] Hai-Ou Li,[1,2,*] Gang Cao,[1,2] Ming Xiao,[1,2] Guang-Can Guo,[1,2] and Guo-Ping Guo[1,2,*]

[1] *CAS Key Laboratory of Quantum Information, University of Science and Technology of China, Hefei, Anhui 230026, China*

[2] *Synergetic Innovation Center of Quantum Information & Quantum Physics, University of Science and Technology of China, Hefei, Anhui 230026, China*

[*]Emails: haiouli@ustc.edu.cn; gpguo@ustc.edu.cn

（Date: 27/11/2018）

## Abstract

Semiconductors, a significant type of material in the information era, are becoming more and more powerful in the field of quantum information. In the last decades, semiconductor quantum computation was investigated thoroughly across the world and developed with a dramatically fast speed. The researches vary from initialization, control and readout of qubits, to the architecture of fault tolerant quantum computing. Here, we first introduce the basic ideas for quantum computing, and then discuss the developments of single- and two-qubit gate control in semiconductor. Till now, the qubit initialization, control and readout can be realized with relatively high fidelity and a programmable two-qubit quantum processor was even demonstrated. However, to further improve the qubit quality and scale it up, there are still some challenges to resolve such as the improvement of readout method, material development and scalable designs. We discuss these issues and introduce the forefronts of progress. Finally, considering the positive trend of the research on semiconductor quantum devices and recent theoretical work on the applications of quantum computation, we anticipate that semiconductor quantum computation may develop fast and will have a huge impact on our lives in the near future.

## Introduction

Recently, the tremendous advances in quantum computation attracted global attention, making this subject again under the spot light since it was first proposed by Richard Feynman [1] in 1982. In the race to build a quantum computer, several competitors emerged like superconducting circuits [2, 3], trapped ions [4, 5], semiconductors [6, 7], nitrogen-vacancy centers [8, 9], nuclear magnetic resonance [10] and etc. Among them, semiconductor is one powerful contender for its significant role in the field of classical computing. They have not only changed our lives with the personal computer, smart phone, Internet and artificial intelligence but also boosted the economic worldwide like the birth of Silicon Valley in the United States of America. With a belief of promoting technology revolution once again in the quantum field, in the last decade, several significant breakthroughs in quantum information processing have been made based on semiconductors. These advances in turn confirm the faith to build a quantum computer out of semiconductor.

Similar to the classical counterpart that is built upon classical bits, a quantum computer is made of quantum bits, which is also called 'qubit'. A qubit is a two level system that exhibits quantum properties: superposition and entanglement. Superposition refers to the ability that a qubit not only can reside in the state $|0\rangle$ or $|1\rangle$ like classical bits, but also can be in the state:

$$|\psi\rangle = \cos(\theta/2)|0\rangle + e^{i\varphi}(\sin\theta)/2\,|1\rangle \qquad (1)$$

Here $\theta$ and $\varphi$ are real numbers that define a point on a unit three-dimensional sphere. Thus an arbitrary qubit state can be described as a point on the surface of a sphere, as depicted in Fig.1 (a), which is termed Bloch sphere. The basis states $|0\rangle$ and $|1\rangle$ are the North and South Pole of the sphere, respectively, while two superposition states $1/\sqrt{2}\,(|0\rangle + |1\rangle)$ and $1/\sqrt{2}\,(|0\rangle - |1\rangle)$ are on the equator. For the property of entanglement, it describes the correlation of different qubits during processing, i.e. a two qubit state can be $1/\sqrt{2}(|01\rangle + |10\rangle)$ that one qubit state depends on the other: if the first qubit were in state $|0\rangle$, the other qubit would be in state $|1\rangle$, and vice versa. By taking advantage of these two significant properties, many quantum algorithms were proposed to give nearly exponential speedup over classical computing for a variety of problems, such as prime factorization [11], data searching [12], numerical optimization [13], chemical simulation [14], machine learning [15] and etc. Since these problems are very common in the fields of banking, internet, business, industry and scientific research, these quantum algorithms are believed to have a widespread use in the future.

All the quantum algorithms are based on a certain quantum computing model, varying from the quantum circuit, one-way quantum computation, adiabatic quantum computation and topological quantum computation. All these four models are equivalent in computational power and among them the quantum circuit model is more frequently used. In the circuit model, it has been proved that arbitrary single-qubit rotations plus two-qubit controlled-NOT gates is universal, i.e. they can provide a set of gates to implement any quantum algorithm [16]. As Fig. 1(a) shows, for a certain initial state $|\psi_i\rangle$ on the Bloch sphere, an arbitrary target state $|\psi_t\rangle$ can be achieved just by successive rotations about z and y axis for $\phi_z$ and $\phi_y$, respectively. In fact, as long as one can control rotations around two different axes of the Bloch sphere, arbitrary single-qubit rotations can be performed, which is also known as universal single qubit control. On the other hand, a two-qubit controlled-NOT (CNOT) gate implies a qubit state can be controlled by another. It acts on two qubits and a π rotation around *x* axis is performed on the target qubit only when the control qubit state is $|1\rangle$. This intriguing phenomenon is shown in Fig. 1(b), an experimental result from Zajac et.al [17], in which the ground state $|0\rangle$ ($|1\rangle$) is denoted by spin-down $|\downarrow\rangle$ (spin-up $|\uparrow\rangle$). In this figure, as manifested by the spin-up probabilities, the left qubit (red) shows rotations around *x* axis as a function of interaction time when the right qubit (blue) is initialized in $|1\rangle$ whereas keeps its initial state all the time when the right qubit is initialized oppositely. The vertical dashed line at which two left qubit states are exactly opposite corresponds to a CNOT gate. Therefore, the core issue of building a quantum computer is to prepare a qubit with high fidelity single- and two-qubit gates. The control fidelity depends on two factors: the coherence time and the manipulation time. Coherence time, also called dephasing time, is usually termed as $T_2$ that indicates how long a qubit can keep its quantum properties, while manipulation time that is characterized by a rotation angle of π ($T_\pi$) or 2π ($T_{2\pi}$) refers to the time required for a single manipulation. In qubit experiments, the coherence time can be obtained by measuring the decay time of Larmor precession and Ramsey fringes [10]. Due to the instrumental imperfections, these decay times are usually smaller than $T_2$ and are termed as $T_2^*$. To get rid of these imperfections, dynamical decoupling pulses can be utilized and the resulting decay time is the intrinsic $T_2$. In some experiments when these two parameters cannot be obtained, the decay time of other coherent oscillations are also used to estimate the qubit coherence, such as the decay time of Rabi oscillations ($T^{Rabi}$). Usually, the Rabi decay time $T^{Rabi}$ is longer than $T_2$ since the concatenation of its oscillations plays a similar role as dynamical decoupling and during half its time the qubit stays at the eigenstate that is less affected by dephasing effect. One application of $T^{Rabi}$ is the proposed quality factor $Q = T^{Rabi}/T_\pi$ [18] to characterize a qubit fidelity. A rough estimate of the qubit fidelity via Q is that a Q ~ 100 suggests a fidelity above 99% and a Q ~ 1000 suggests a fidelity above 99.9%. In the fields lacking fidelity measurements, the Q value is often used as a reference. Nevertheless, no matter how high the fidelity is, the small errors can still be propagated and amplified through successive manipulations until the computation process is destroyed. To tackle this problem, a solution is to build a fault tolerant quantum computer with qubits encoded by error correcting codes. An example of these codes is the surface code, which requires 2-dimensional array of qubits with single- and two- qubit gate fidelities above the threshold of 99% [19]. If qubits can be prepared meeting this requirement, millions of qubits encoded by surface code can be employed for running effective quantum algorithms.

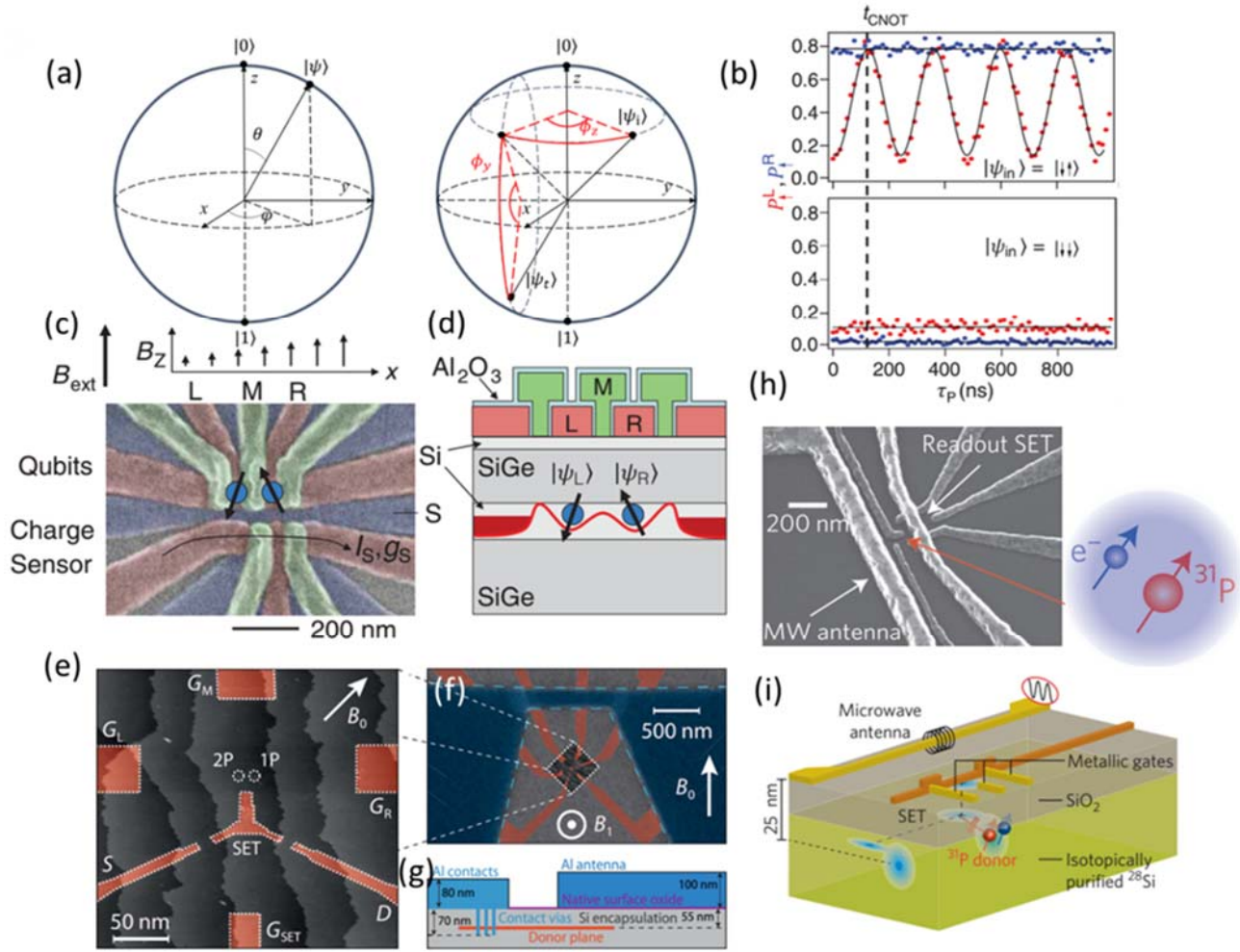

**Figure 1.** Single- and two- qubit gate control and devices for semiconductor qubits. (a) Bloch sphere representation of a qubit. A superposition state $|\psi\rangle$ can be represented by a point on the sphere (left). An arbitrary rotation from initial state $|\psi_i\rangle$ to the target state $|\psi_t\rangle$ can be decomposed by successive rotations about z axis and y axis for $\phi_z$ and $\phi_y$, respectively (right). (b) The spin-up probability of the spin-up state for the right qubit $P_\uparrow^R$ (blue) and the left qubit $P_\uparrow^L$ (red) as a function of interaction time $\tau_P$ for input states $|\downarrow\uparrow\rangle$ and $|\downarrow\downarrow\rangle$. The vertical dashed line at $\tau_P = $ 130 ns corresponds to a CNOT gate (Adapted from ref. [17]) . (c) and (d) are false-color SEM image and a schematic cross-section of a Si/SiGe DQD, respectively. The DQD with two electrons confined in the potential created by gates L, M and R is used to form two spin-1/2 qubits and a SET under the DQD is used to work as a charge sensor. A slanting Zeeman field was created by a micro-magnet (not shown) for qubit control (Adapted from ref. [17]). (e), (f) and (g) are images and schematics for the device fabricated by STM hydrogen lithography (Adapted from ref. [30]).

(e) Large-scale STM image of the device, red areas are P-doped to form a SET, source and drain leads, and electrostatic gates. A donor molecule (2P) and single donor (1P) are shown by two circles. (f) False color composite SEM and STM image showing the buried donor structures (red) and the aluminum antenna (blue). (g) Vertical cross-section of the donor device, showing the thickness (not to scale) and relative position of the silicon, phosphorus, oxide and aluminum layers. (h) and (i) are SEM image and schematic oblique view of a device fabricated by ion implantation, highlighting the position of the P donor, the MW antenna and the readout SET (Adapted from ref. [31] and [51]).

In 1998, Loss and DiVincenzo [20] first proposed to utilize semiconductor quantum dots for manipulating single spins as qubits. A typical device of gate-defined lateral quantum dots is shown in Fig. 1(c) and (d), the electrodes on the surface of the Si/SiGe heterostructure can form quantum potentials in the Si well to trap electrons, and the electron spins can be manipulated as qubits when an external magnetic field is applied. The upper half of Fig.1 (c) is a double quantum dot (DQD) to form two spin-1/2 qubits and the lower half is a single quantum dot (SQD) acting as a charge sensor to measure the charge states of the DQD, which is also called a single electron transistor (SET). In fact, quantum dots can be formed in various systems, including GaAs/AlGaAs heterostructures [21], the silicon Metal-Oxide-Semiconductor (MOS) and silicon on insulator (SOI) [22], nanowires [23], nanotubes [24], graphene [25], van der Waals heterostructures [26, 27], and self-assembled crystals[28]. It's worth mentioning that quantum dots based on $Si/SiO_2$ and SOI technology are both CMOS compatible and in this article we denote the former as silicon MOS and the latter as SOI for clarity. On the heels of the proposal for quantum dot based electron spins, Bruce Kane [29] presented that the nuclear spin of a single $^{31}$P donor in silicon can also be controlled as a qubit. There are two approaches to fabricate the device: scanning tunneling-microscopy (STM) hydrogen lithography and ion implantation. For the former approach, the STM tip enables atomic-scale precision of placing P atoms in a silicon. Fig. 1(e) is a STM image of a device fabricated using this approach, showing a single donor (1P) and a donor molecule (2P) in the center for spin manipulation and beneath them is a SET for charge sensing. The blue area in Fig. 1(f) is an aluminum antenna generating an oscillating magnetic field over the device, and Fig. 1(g) is the vertical cross-section showing the relative position of the antenna and the silicon device [30]. For the latter approach, P ions are implanted into a very small region of the silicon using mask resists. Fig.1 (g) and (h) are the scanning electron microscope (SEM) image and the schematic of a device fabricated by ion implantation. In Fig. 1(g), a P donor was implanted in the area denoted by the red arrow, and spins of both the electron bound to the donor and the donor nucleus can be used as qubits [31]. Also, the SET and the Al antenna are used for readout and manipulation. In 2003, Hayashi and co-workers also investigated the coherent manipulation of electronic states of a DQD in the GaAs/AlGaAs heterostructure and showed the opportunity to implement a charge qubit [32] in a semiconductor DQD. These proposals together resulted in a subsequent firestorm of experimental activities [7]. Till now, single- and two-qubit gate control have been achieved with fidelity above 99.9% [18, 33] and 98% [34] respectively, approaching the surface code threshold for fault tolerant computing. Also, thanks to the advanced semiconductor technology, several proposals taking advantage of today's semiconductor processing tools to scale up to 2D grids [35-39] have been put forward. Therefore, it is believed that there is a huge opportunity to realize a scalable fault tolerant semiconductor quantum computer in the future.

In the following, we will begin with discussing single qubit control for different types of semiconductor qubits and then move to two qubit gates. Then, the challenges and also the opportunities for building a quantum computer will be discussed. In the end, we will introduce the views of semiconductor quantum computation around the world and anticipate that the research on semiconductor quantum devices may have a great influence in the following years.

# Single qubit gate in semiconductor

As discussed in the introduction, both the spin and charge degrees of electrons and donor nucleus can be employed as qubits [7, 40]. For the spin degree, spin-1/2 qubits, singlet-triplet qubits and exchange-only qubits were proposed and realized in experiments successively. And to take advantage of both spin and charge degrees, the hybrid qubit was also presented as a competitive candidate. In this section, we restrict our scope to the single qubit control of those qubits, and will discuss the two qubit gate in the next section.

# Spin-1/2 qubit

Once an electron or nucleus is put into magnetic field $B_0$, the energy levels of spin-up and spin-down are no longer degenerate and split by so-called Zeeman energy. This is a two level system that can be used as a qubit and we name it spin-1/2 qubit to distinguish it from other types of spin qubits. To manipulate this type of qubit, microwave (MW) bursts via an antenna were used to generate oscillating magnetic field [30, 31], as illustrated in Fig.1 (f) and (h). This approach is called electron spin resonance (ESR) for controlling electron spins or nuclear magnetic resonance (NMR) for controlling nuclear spins. In the rotating frame, the control Hamiltonian can be written as:

$$\boldsymbol{H}_R \approx (-\omega_0 + \omega)\sigma_z/2 - \omega_R[\cos(\phi)\sigma_x/2 - \sin\phi\,\sigma_y/2] \tag{2}$$

For simplicity, we use the natural unit throughout the article. Here, $\omega$, $\omega_0$, $\omega_R$ are MW frequency, Larmor frequency and Rabi frequency, respectively. The latter two satisfy the condition $\omega_0 = \Upsilon|\boldsymbol{B}_0|$ and $\omega_R = \Upsilon|\boldsymbol{B}_1|$ with $\Upsilon$ the gyromagnetic ratio, $\boldsymbol{B}_0$ the external magnetic field and $\boldsymbol{B}_1$ the oscillating magnetic field perpendicular to $\boldsymbol{B}_0$. Thus the Larmor frequency corresponds to the Zeeman energy splitting of the qubit. $\boldsymbol{B}_1$ is produced by the MW antenna and its magnitude is directly proportional to the current through the antenna. The Pauli matrices $\sigma_z$, $\sigma_x$ and $\sigma_y$ suggest rotations around the *z* axis, *x* axis and *y* axis of the Bloch sphere, respectively. Therefore, a sequence of MW bursts with a frequency $\omega = \omega_0$ on resonance with the qubit and initial phase $\phi = 0$ will drive the qubit rotating around *x* axis. Specially, the nutation between $|0\rangle$ and $|1\rangle$ is usually called Rabi oscillation. When the MW is halted for a time, or the relative phase of successive MW bursts is varied, the qubit will acquire a rotation angle around *z* axis. The universal single spin control thus can be achieved using this approach. Alternatively, another approach to manipulate spin-1/2 qubit is electric-dipole spin resonance (EDSR). In this approach, a magnetic field gradient is applied with the help of spin-orbital coupling (SOC) of the semiconductor or an integrated micro-magnet, and the electron in this environment can feel an effective oscillating magnetic field if it is driven by an oscillating electric field. Therefore, MW bursts can be applied directly on a single electrode and $\boldsymbol{B}_1$ is proportional to its voltage amplitude. One example using this approach is shown in Fig. 1(c), there is a magnetic field gradient in the device generated by an integrated micro-magnet (not shown), and the MW bursts are applied on gate S for qubit control [17].

Readout of the spin-1/2 qubits relies on a spin-charge conversion as spin-selective tunneling [41-43] or spin blockade [44, 45], and after the conversion the charge signal is detected by a nearby charge sensor. The procedure for spin selective tunneling is illustrated in Fig. 2(a) and (b), when a spin-1/2 qubit is under MW control, the energy levels of both spin states are under the Fermi level of the drain, and after control, the energy levels in the quantum dot are tuned so that the energy level of spin-up is higher than the Fermi level of the drain and spin-down is lower. In this energy level alignment, only the electron with spin-up can tunnel out of the quantum dot and thus the spin state can be distinguished by observing the electron tunneling signal. This approach was first demonstrated by Elzerman et al. in 2004 [41], and they achieved single-shot readout of a single electron spin for the first time. An adaption of Elzerman's method is to use the tunneling rate difference instead of energy difference of two energy levels to differentiate spin states, which was first demonstrated by Hanson et al. in 2005 [42]. Here we term the former energy-selective readout and the latter tunnel-rate selective readout. Inspired by Elzerman et al.'s work, in 2010, Morello et al. demonstrated the first single-shot spin readout of an electron bound to a donor in silicon, in which they used the electrochemical potential of the SET to play the role of the Fermi level of a drain for energy selection [43]. As for the spin blockade, it utilizes another spin as an ancilla qubit to read the spin state in the singlet-triplet basis. There are four basis states for two spins in a magnetic field and can be sorted into a singlet and three triplets:

$$S = 1/\sqrt{2}(|\uparrow\downarrow\rangle - |\downarrow\uparrow\rangle), T_0 = 1/\sqrt{2}(|\uparrow\downarrow\rangle + |\downarrow\uparrow\rangle), T_+ = |\uparrow\uparrow\rangle, T_- = |\downarrow\downarrow\rangle \quad (3)$$

Here, $S$ and $T_0$ are separated with exchange interaction strength $J$, and the three triplets are split by Zeeman energy $E_z$. We denote the singlet with each electron occupying one quantum dot as $S(1, 1)$ and the one with two electrons both occupying the right dot as $S(0, 2)$. This type of denotation also applies to the triplets. If we suppose both spins are initialized in $|\downarrow\downarrow\rangle$, the MW bursts on the left spin will lead the two spin state to oscillate between $|\downarrow\downarrow\rangle$ and $|\uparrow\downarrow\rangle$. In experiments, $|\downarrow\downarrow\rangle$ is usually mapped to $T_-(1, 1)$ and $|\uparrow\downarrow\rangle$ is mapped to $S(1, 1)$. As illustrated in Fig. 2(c) and (d), only the $S(1, 1)$ state can transit to $S(0, 2)$ and other states are prohibited because of spin blockade. Thus a nearby charge sensor that can differentiate charge state (1, 1) and (0, 2) is able to read out the spin state. For simplicity, in the figure we denote $T(1, 1)$ for those triplets. This measurement method was first demonstrated by Petta et al. in 2005, and with mapping $|\uparrow\downarrow\rangle$ to $S$ and $|\downarrow\uparrow\rangle$ to $T_0$, they implemented a controlled two qubit gate exchanging the spin direction [44]. Then combined with MW bursts, Koppens et al. demonstrated the first driven coherent single spin rotations in 2006 [45]. Furthermore, the visibility of readout in the singlet-triplet basis can be enhanced by charge state latching [46, 47] and intermediate excited states [48]. In Harvey-Collard et al.'s work, they achieved a measurement fidelity as high as 99.86% via charge state latching, and in 2018, Fogarty et al. demonstrated measuring a spin-1/2 qubit using this new method [49].

Since Petta et al. and Koppens et al. first demonstrated a two- and single- qubit gate respectively in GaAs quantum dots, they met a serious problem that the hyperfine interactions (HI) with the GaAs host nuclei has a nontrivial influence on the coherence of the spin-1/2 qubit and limits its dephasing time $T_2^*$ only to tens of nanoseconds [50]. An alternative approach is to use group IV host materials, Si or Ge, to eliminate the random nuclear spins [51]. In 2014, Kawakami et al. demonstrated a spin-1/2 qubit with $T_2^* \sim 1$ μs in natural silicon [52]. Still, the residual $^{29}$Si nuclear spins in natural silicon can deteriorate the dephasing time, and to overcome this, Veldhorst et al. performed a similar experiment in purified silicon with $^{29}$Si down to 800 ppm. The great reduction of $^{29}$Si enabled them to measure a $T_2^*$ as long as 120 μs and obtained a fault tolerant fidelity of 99.6% [53], which is at the threshold of surface code for fault tolerant quantum computing. However, limited by heating of ESR, the manipulation time of their device is up to $T_\pi \sim 1$ μs. In 2017, by using EDSR rather than ESR, Yoneda et al. reduced the manipulation time $T_\pi$ down to ~17 ns and with keeping $T_2^* \sim 20$ μs, they reported a control fidelity over 99.9%. The high-frequency Rabi oscillation with $T_\pi \sim 128$ ns and the randomized benchmarking of the control fidelity for different types of gates are shown in Fig. 2(e) and (f), respectively. As for donors in silicon, in 2014, Muhonen et al. demonstrated a dephasing time as long as 600 ms for $^{31}$P$^+$ spin and reached a control fidelity >99.99%. Moreover, the intrinsic dephasing time measured by dynamical decoupling pulses resulted in a $T_2$ exceeding 30 seconds, implying its future application in storing quantum information. For the electron spin bound to the donor, they also obtained a $T_2^* \sim 270$ μs and a fidelity over 99% [31].

Except for electrons and nuclei, hole spins can also be encoded as spin-1/2 qubits. The small HI and strong SOC of the hole spin promises both a long dephasing time and a short manipulation time, therefore, this field becomes very active since the first measurement of the coherence time of a single hole spin in a InGaAs quantum dot [54]. Until now, the hole spin qubits have been demonstrated in a SOI DQD [55] and a Ge hut wire DQD [56]. With strong SOC, the best manipulation rate can be over 140 MHz, i.e. the rotation time $T_{2\pi}$ can be as short as ~7 ns. However, the measured dephasing times were both at least one order of magnitude lower than that of electron spins, and the qubit control cannot be proved at a single hole level. Other approaches include manipulating spins of holes that are bound to acceptors in silicon [57] or that are trapped in quantum dots fabricated from the p-GaAs/AlGaAs heterostructure [58], the silicon MOS [59], Ge/GeSi heterostructures [60] and core shell nanowires [61].

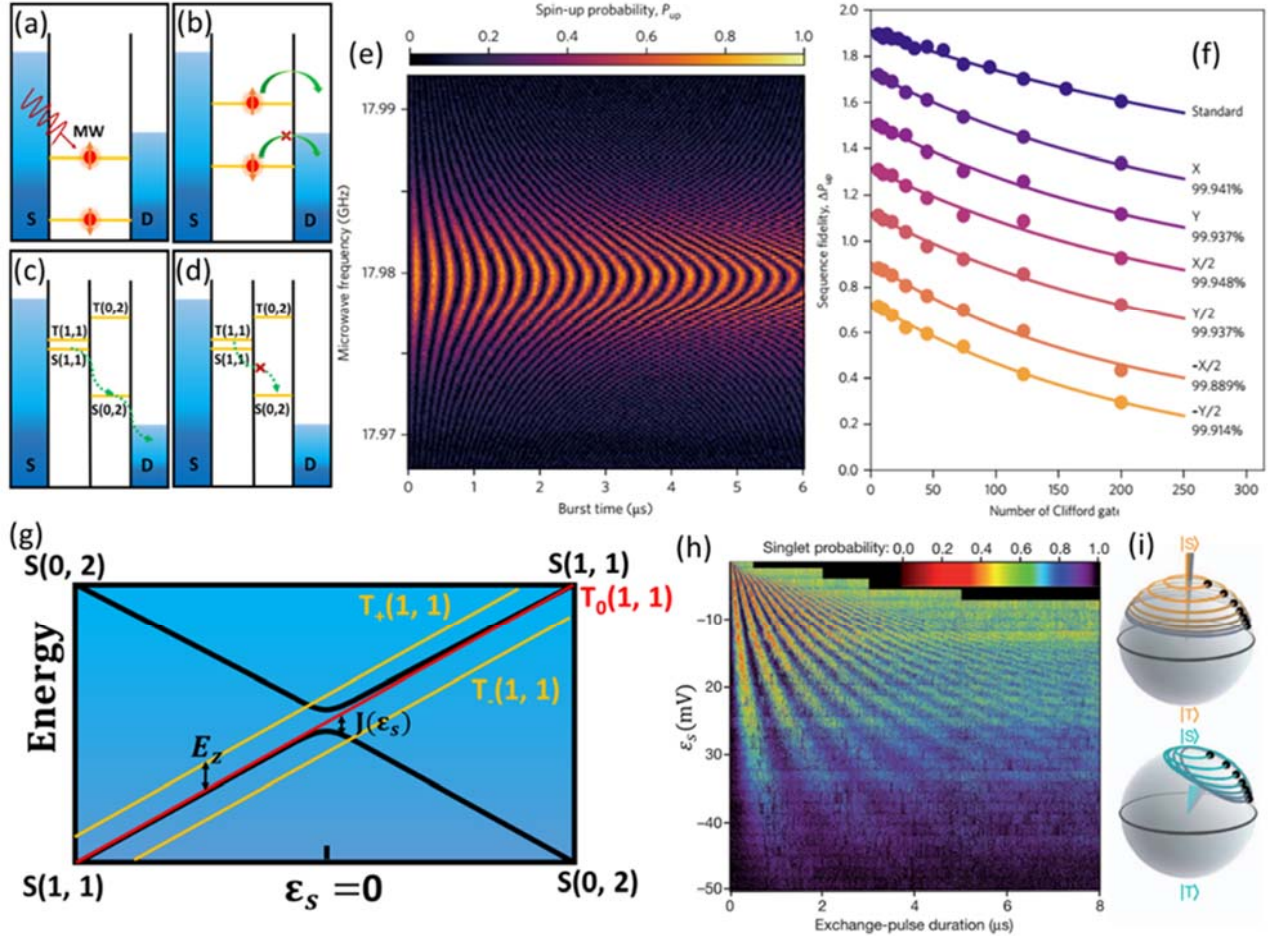

**Figure 2.** Spin-1/2 qubit and singlet-triplet qubit. (a)-(b) are diagrams showing the process of control and readout based on spin selective tunneling. (a) At the stage for qubit control, both energy levels of spin-up and spin-down are under the Fermi level of drain. (b) At the stage for readout, the energy levels in the dot are tuned so that the Fermi level of drain is between the energy levels of spin-down and spin-up. (c) and (d) are diagrams showing the phenomenon of spin blockade: $S(1, 1)$ can move to $S(0, 2)$ while $T(1, 1)$ cannot. (e) The probability of spin-up $P_{up}$ as a function of MW burst time and frequency detuning (Adapted from ref. [17, 18]). (f) Sequence fidelities for standard (top-most) and interleaved randomized benchmarking (annotated in the figure along with extracted fidelities). Traces are offset by an increment of 0.2 for clarity. Visibilities are within $0.72 \pm 0.012$ (Adapted from ref. [17, 18]). (g) Energy level spectrum of two spin states in a DQD as functions of detuning $\varepsilon_s$. A magnetic field splits the triplet states by Zeeman energy $E_z$ and the exchange interaction splits $S$ and $T_0$ by $J(\varepsilon_s)$. (h) Singlet probability as a function of exchange-pulse duration and detuning $\varepsilon_s$ (Adapted from ref. [68]). (i) Bloch sphere representations of state evolution in the case $J(\varepsilon_s) > \Delta E_z$ (top) and $J(\varepsilon_s) < \Delta E_z$ (bottom) (Adapted from ref. [68]).

## Singlet-triplet qubit

Another type of spin qubit is encoded by two eigenstates of two spins [62]. Usually, the encoded states are $S$ and $T_0$, and we thus call it singlet-triplet qubit. The effective control Hamiltonian can be written as follows,

$$\boldsymbol{H_{ST}} = J(\varepsilon_s)\sigma_z/2 + \Delta \boldsymbol{E_z}\sigma_x/2 \tag{4}$$

Here, $J(\varepsilon_s)$ is the energy of exchange splitting of $S$ and $T_0$, where the detuning $\varepsilon_s$ denotes the electrochemical potential difference of different charge occupation states. And $\Delta \boldsymbol{E_z}$ is the Zeeman energy difference of two spins, which may be caused by different g-factor [63], i.e. $\Delta \boldsymbol{E_z} = \Delta g \mu_B \boldsymbol{B_z}$, or magnetic field gradient [64, 65], i.e. $\Delta \boldsymbol{E_z} = g\mu_B \Delta \boldsymbol{B_z}$. As shown in Fig. 2(g), when the detuning point is set negative and far away from zero, $J(\varepsilon_s)$ will be vanishing and thus the qubit will rotate around $x$ axis; on the contrary, when the detuning point is tuned in the positive direction until $J(\varepsilon_s) \gg \Delta \boldsymbol{E_z}$, the qubit will rotate around $z$ axis. In this control procedure, only the parameter $\varepsilon_s$ is used and thus the need of ESR or EDSR is removed compared to spin-1/2 qubits. After manipulation, the qubit can be measured directly using spin blockade. The singlet-triplet qubit was first demonstrated experimentally in GaAs quantum dots with a $T_2^* \sim$ 10 ns and a rotation period $T_{2\pi} \sim$ 720 ps around $z$ axis by Petta et al. in 2005 [44]. The dephasing time is mainly limited by the fluctuating nuclear field and many following researches concentrated on how to improve it. By using dynamical nuclear polarization (DNP) for controlling nuclear field in a feedback loop, in 2010, Bluhm et al. demonstrated a $T_2^* \sim$ 94 ns [66], and further with dynamical decoupling pulses to filter low-frequency noises, they achieved an intrinsic dephasing time $T_2$ exceeding 200 μs [67]. Similar to spin-1/2 qubits, silicon was also expected to replace GaAs as a host material for singlet-triplet qubits. In 2012, Maune et al. first demonstrated a singlet-triplet qubit in a Si/SiGe DQD, reporting a $T_2^* \sim$ 360 ns [68]. The $S$-$T_0$ oscillations in this experiment can be observed in Fig. 2(h). Fig. 2(i) are the Bloch sphere representations of state evolution in the case $J(\varepsilon_s) > \Delta \boldsymbol{E_z}$ and $J(\varepsilon_s) < \Delta \boldsymbol{E_z}$. However, spin blockade can be lifted easily due to the small splitting of two low lying valley states in Si/SiGe heterostructure, which puts a great hurdle in front of the reproducibility of singlet-triplet qubits in this material. In 2014, Shulman et al. found real-time Hamiltonian estimation (RHE) could be used to suppress qubit dephasing and they measured a $T_2^*$ more than 2 μs in a GaAs DQD, even one order of magnitude over that in silicon [69]. In 2016, Malinowski et al. improved the design of dynamical decoupling sequences to use it as a notch filter for nuclear noises and improved $T_2$ to 870 μs [70]. As for charge noises, in 2017, Nichol et al. [71] discovered that a large $\Delta \boldsymbol{E_z}$ could suppress charge noises and combined it with RHE, they reported a record single-qubit gate fidelity of ~99% in GaAs.

Apart from these, singlet-triplet qubits can also be encoded by $S$ and $T_+$ [72], or implemented in other systems, such as donors in silicon [73, 74] and hybrid donor-dot architecture [75]. It is noteworthy that for donors in silicon, the transition between (1, 1) and (0, 2) are harder to distinguish for the special charge sensor arrangement and thus the energy-selective readout or tunnel-rate selective readout like spin-1/2 qubits are preferred. These two readout methods for singlet-triplet qubits have been investigated by Broome et al. [74] and Dehollain et al. [73], respectively. And for the hybrid donor-dot singlet-triplet qubits, the readout relies on the aforementioned latching enhanced spin blockade.

## Exchange-only qubit

Though singlet-triplet qubits can be driven all electrically, it still needs a Zeeman energy difference to achieve universal single qubit control. How about implementing a qubit solely by exchange interaction? This idea leads to the exchange-only qubit [76]. As illustrated in Fig. 3(a), this type of qubit is composed of three electrons in a triple quantum dot (TQD) [77-79]. There are eight basis states for three spins, and among them

$|S_l\rangle = 1/\sqrt{2}(|\uparrow\downarrow\uparrow\rangle - |\downarrow\uparrow\uparrow\rangle)$ and $|T_l\rangle = 1/\sqrt{6}(|\downarrow\uparrow\uparrow\rangle + |\uparrow\downarrow\uparrow\rangle - 2|\uparrow\uparrow\downarrow\rangle)$ are separated by exchange splitting $J_l(\varepsilon_x)$, $|S_r\rangle = 1/\sqrt{2}(|\uparrow\uparrow\downarrow\rangle - |\uparrow\downarrow\uparrow\rangle)$ and $|T_r\rangle = 1/\sqrt{6}(|\uparrow\uparrow\downarrow\rangle + |\uparrow\downarrow\uparrow\rangle - 2|\downarrow\uparrow\uparrow\rangle)$ are separated by $J_r(\varepsilon_x)$. Here, $|S_l\rangle$ ($|S_r\rangle$) and $|T_l\rangle$ ($T_r$) are singlet-like state and triplet like states, respectively, which can be inferred from the state of the left (right) two spins. The two exchange splitting energy $J_l(\varepsilon_x)$ and $J_r(\varepsilon_x)$ are associated with the left pair and the right pair of quantum dots, respectively, and the detuning $\varepsilon_x$ denotes the relative electrochemical potential of the charge configuration (2, 0, 1), (1, 1, 1) and (1, 0, 2). As depicted in Fig. 3(a), the ground state $|0\rangle = 1/\sqrt{6}(|\uparrow\uparrow\downarrow\rangle + |\downarrow\uparrow\uparrow\rangle - 2|\uparrow\downarrow\uparrow\rangle)$ and the excited state $|1\rangle = 1/\sqrt{2}(|\uparrow\uparrow\downarrow\rangle - |\downarrow\uparrow\uparrow\rangle)$ are encoded in the center of (1, 1, 1) charge configuration with $J_l(\varepsilon_x) = J_r(\varepsilon_x)$. Also, there are two extra states $|Q\rangle = 1/\sqrt{3}(|\uparrow\uparrow\downarrow\rangle + |\uparrow\downarrow\uparrow\rangle - 2|\downarrow\uparrow\uparrow\rangle)$ and $|Q_+\rangle = |\uparrow\uparrow\uparrow\rangle$ in the energy level spectrum that may offer leakage channels when the qubit is under control. The control Hamiltonian can be described as:

$$H_{EX} = -J_l(\varepsilon_x)\sigma_l/2 - J_r(\varepsilon_x)\sigma_r/2 \qquad (5)$$

in which,

$$\sigma_l = (\sigma_z - \sqrt{3}\sigma_x)/2, \sigma_r = (\sigma_z + \sqrt{3}\sigma_x)/2 \qquad (6)$$

In the Bloch sphere, the axes $\sigma_l$ and $\sigma_r$ are 120° apart and thus the universal single qubit control can be achieved by directly tuning $J_l(\varepsilon_x)$ and $J_r(\varepsilon_x)$ via detuning pulses. This method is called Larmor precession and as an example, a control pulse sequence (yellow) is drawn in Fig. 3(b) indicating a detuning pulse from $|S_l\rangle$ to (1, 1, 1). After manipulation, the qubit state can be measured via spin blockade with $|0\rangle$ mapped to $|S_l\rangle$ ($|S_r\rangle$) and $|1\rangle$ mapped to $|T_l\rangle$ ($|T_r\rangle$). In 2010, Laird et al. first demonstrated an exchange-only qubit in a GaAs TQD with this approach [77], and then, in 2013, Medford et al. measured a inhomogeneous dephasing time $T_2^* \sim 25$ ns and a rotation time $T_{2\pi}$ as short as ~21 ps [78]. The coherent oscillations in their experiment are shown in Fig. 3(c). Another approach to control exchange-only qubits is to use Rabi oscillations. As the red pulse sequence in Fig. 3(b) shows, qubit manipulation can be implemented directly by applying MW bursts at zero detuning with a frequency in resonance with the energy gap between $|0\rangle$ and $|1\rangle$. The qubit controlled in this way is also called resonant exchange qubit. In 2013, Medford et al. demonstrated a resonant exchange qubit and reported an intrinsic dephasing time $T_2 \sim 19$ μs and a rotation time $T_{2\pi} \sim 10$ ns [79]. To make further improvements, other investigations also include reducing magnetic noise by performing experiments in silicon quantum dots [80] and suppressing charge noise by using MW bursts in a highly symmetric regime [81]. Moreover, the spin states of a TQD can also be controlled through Landau-Zener-Stückelberg (LZS) interferences. This was demonstrated by Gaudreau et al. in 2011[82] and they encoded a qubit using the state $|0\rangle$ and $|Q_+\rangle$. The hyperfine interaction that couples these two states results in two anti-crossings in the energy level spectrum, which are denoted by dotted circles in Fig. 3(b). An adiabatic pulse passing through one of the anti-crossings with an appropriate rise time can create a superposition state of $|0\rangle$ and $|Q_+\rangle$ due to Landau–Zener transition, and after a time the pulse goes across the anti-crossing again and back to its original position, resulting in LZS interferences. After that, the measurement in the basis of qubit eigenstates will show corresponding coherent oscillations. From the fit to the LZS model with their experimental results, they extracted a dephasing time $T_2^*$ around 8-15 ns.

## Charge qubit

Besides the spin degree, quantum control of the charge states of an electron is also of interest. For a charge qubit, the ground state $|0\rangle$ and the excited state $|1\rangle$ can be defined by the excess electron occupation of a DQD, and as illustrated in Fig. 3(d), they are usually denoted by $|R\rangle$ and $|L\rangle$, respectively. Readout of the qubit states can be implemented directly by a proximate charge sensor, a SET or a quantum point contact (QPC), or just the transport current from source to drain, so that it removes the need of any conversion like spin qubits. In Fig.

3(d), the current through QPC is shown by the white arrow. The energy levels are depicted in Fig. 3(e), with a Hamiltonian [32, 83, 84]:

$$H_C = \varepsilon_c \sigma_z/2 + t\sigma_x \tag{7}$$

Here $\varepsilon_c$ denotes the detuning energy between $|L\rangle$ and $|R\rangle$, and $t$ is the inter-dot tunneling rate. As shown in Fig. 3(e), a rectangular nonadiabatic voltage pulse (orange) that drives the qubit from ground state ($\varepsilon_c > 0$) to the anti-crossing ($\varepsilon_c = 0$) can induce Larmor precession, resulting in a rotation around *x* axis. If $\varepsilon_c$ is kept still at the ground state, which is far away from the anti-crossing, the inter-dot tunneling rate will vanish, leading to a rotation around *z* axis. Using this approach, the charge qubit was first demonstrated in GaAs quantum dots by Hayashi et al. in 2003, reporting a coherence time $T_2^* \sim$ 1 ns and a rotation time $T_{2\pi} \sim$ 435 ps [32]. Then the inhomogeneous dephasing time $T_2^*$ was determined by Ramsey fringes as 60 ps [85]. Experiments based on Si/SiGe quantum DQDs were also reported with $T_2^* \sim$ 2.1 ns measured by Larmor oscillations, $T_2^* \sim$ 127 ps and $T_2 \sim$ 760 ps obtained by the Ramesy fringes and dynamical decoupling pulses, respectively [86]. Here, the flat band at the anti-crossing point makes the qubit less affected by the charge noise induced by detuning compared to the steeper point at $\varepsilon_c > 0$ and thus the coherence time of rotations around *x*-axis (Larmor oscillations) is much longer than that of *z*-axis (Ramesy fringes and dynamical decoupling pulses). Since the LZS interference is less sensitive to certain types of noise, the charge qubit was also investigated through LZS interferences. In 2012, Stehlik et al. performed LZS interferometry of a semiconductor charge qubit via continuous microwave irradiation and observed the coherent oscillations of the qubit states [87]. In 2013, Cao et al. first observed the LZS interferences in the time domain and demonstrated an ultrafast universal qubit control with $T_{2\pi}$ as short as ~10 ps and intrinsic dephasing time $T_2$ up to 4 ns that was extracted from the amplitude spectroscopy [88]. The adiabatic short pulse they used to drive the qubit is shown in Fig. 3(e), as described in the previous part, the LZS interference is finished after the pulse goes across the anti-crossing and back to its original position. The measured interferences as a function of pulse amplitude *A* are depicted in Fig. 3(f), and as shown by the Bloch spheres labelled at two interference nodes, the qubit is rotated around *z* axis by $2\pi$ between every two successive interference fringes while the rotation angle of *x* axis, *θ,* increases monotonically with pulse amplitude. Therefore, the qubit can be rotated around both *x* and *z* axis within a single pulse and these rotations can be controlled arbitrarily by adjusting the pulse amplitude. Moreover, the charge qubit can also be controlled by applying resonant MW bursts at $\varepsilon_c = 0$ to induce Rabi oscillations and the two-axis control using MW bursts is just like that of spin-1/2 qubits and resonant exchange qubits. Here, the ground state and the excited state are changed to $|0\rangle \approx (|L\rangle + |R\rangle)/\sqrt{2}$ and $|1\rangle \approx (|L\rangle - |R\rangle)/\sqrt{2}$. In 2015, Kim et al. implemented a resonant charge qubit in a Si/SiGe quantum DQD, reporting a $T_2^*$ of 1.3 ns and $T_2 \sim$ 2.2 ns [89]. With the improvement in coherence time, they measured an average qubit fidelity greater than 86%. In 2018, a research on valley-orbit states in silicon also implied that the hybridized valley-orbit states can potentially be employed for higher fidelity control, where the energy band is flat with respect to a larger range of detuning [90].

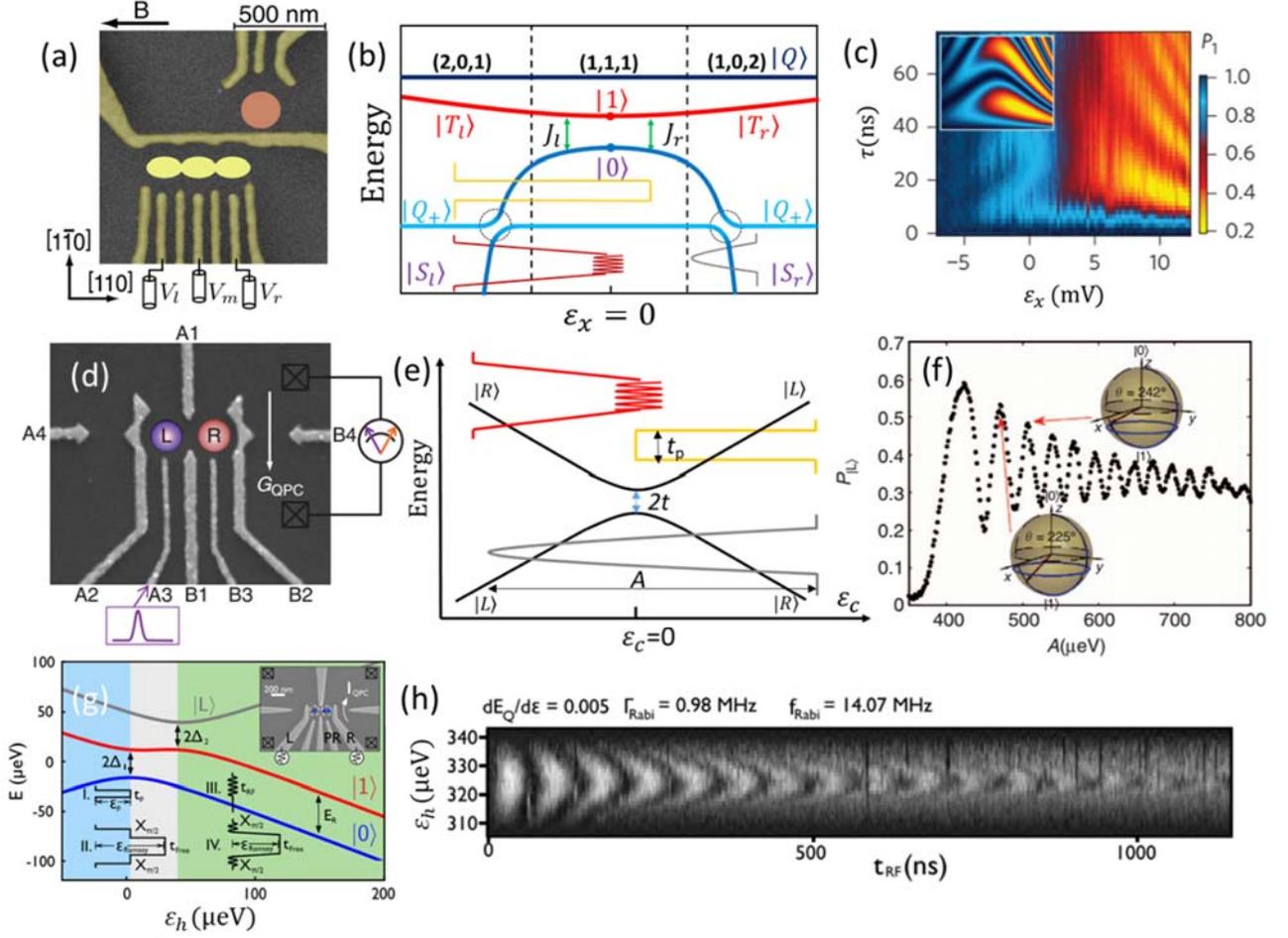

**Figure 3.** Implementations of the exchange-only qubit, the charge qubit and the hybrid qubit. (a) False colored SEM image of a TQD device for an exchange-only qubit, with a SET on the top for charge sensing (Adapted from ref.[78]). (b) Energy levels as a function of detuning $\varepsilon_x$ for the exchange-only qubit. Two anti-crossing are shown by two dotted circles. Yellow, red and grey pulse shapes are shown to denote the relative positions of detuning for Larmor oscillation, Rabi oscillation and LZS interferences. (c) The probability $P_1$ of detecting the state $|S_l\rangle$ as a function of pulse position $\varepsilon_x$ and wait time τ. Inset is a simulation result of qubit evolution as a function of exchange without noise (Adapted from ref. [78]). (d) A SEM image of a DQD device for a charge qubit with two QPCs for readout. $|L\rangle$ and $|R\rangle$ are denoted by two circles in the DQD (Adapted from ref. [88]). (e) Energy levels of a charge qubit as a function of detuning $\varepsilon_c$. Red, yellow and grey pulse shapes are shown to denote the relative positions of detuning for Rabi oscillation, Larmor oscillations and LZS interferences. (f) Charge state probability $P_{|L\rangle}$ as a function of LZS pulse amplitude A. The Bloch sphere representations for two interference nodes are also labelled (Adapted from ref. [88]). (g) Energy spectrum as a function of $\varepsilon_h$ and pulse sequences for the hybrid qubit. Inset is a SEM image of a DQD device with a QPC for readout (Adapted from ref. [94]). (h) Rabi oscillations demonstrating a decay time longer than 1 μs (Adapted from ref. [94]).

# Hybrid qubit

Inspired by the fact that the coherence times of spin qubits are usually very long and the manipulation times of charge qubits are very short, one may question whether we can create a new type of qubit combining advantages of both. An attempt originates from this idea is the hybrid qubit [91, 92]. This type of qubit is encoded by two eigenstates of three electron spins in a DQD and was first demonstrated in a Si/SiGe heterostructure [93]. Fig. 3(g) shows its energy levels as well as the device setup [94]. The two lowest energy levels for qubit control are

$$|0\rangle = |\downarrow\rangle_L |S\rangle_R, \quad |1\rangle = 1/\sqrt{3}|\downarrow\rangle_L |T_0\rangle_R - \sqrt{2/3}|\uparrow\rangle_L |T_-\rangle_R \quad (8)$$

The subscript $L$ ($R$) denotes the spin state in the left (right) quantum dot, and the higher state $|L\rangle$ in Fig. 3(g) is a primary leakage channel. In the basis of these three states, the Hamiltonian can be written as

$$\boldsymbol{H}_C = \begin{pmatrix} -\varepsilon_h/2 & t_1 & t_2 \\ t_1 & -\varepsilon_h/2 & 0 \\ t_2 & 0 & -\varepsilon_h/2 + E_R \end{pmatrix} \quad (9)$$

Here, $t_1$ and $t_2$ are the tunnel couplings between $|0\rangle$ and $|1\rangle$, $|1\rangle$ and $|L\rangle$, respectively. And $\varepsilon_h$ is the detuning between charge state (2, 1) and (1, 2), while $E_R$ is the energy separation between the two lowest valley-orbit states in the right dot. The energy level spectrum can be divided into three regions: charge-like region (blue), hybrid region (grey) and spin-like region (green). In the spin-like region, $E_R$ is just the splitting energy of $|0\rangle$ and $|1\rangle$. The charge-like region can be used for readout using spin blockade with $|S\rangle_R$ mapped to $|0\rangle$ and $|T_0\rangle_R$ as well as $|T_-\rangle_R$ mapped to $|1\rangle$. In the readout regime, spin blockade will permit the transition between (1, 2) and (2, 1) for $|S\rangle_R$ and prohibit it for $|T_0\rangle_R$ and $|T_-\rangle_R$. Therefore, $|0\rangle$ and $|1\rangle$ can be distinguished from the charge occupation after the conversion. For qubit control, as shown by the pulses labelled as (I)-(IV) in Fig. 3(g), it can be performed either in the hybrid region by Larmor precession or in the spin-like region by Rabi oscillation. In the Larmor precession regime, a control pulse stops at $\varepsilon_h = 0$ and $\varepsilon_h > 0$ will rotate the qubit about $x$ axis and $z$ axis, respectively. In 2014, Kim et al. measured a control fidelity of 85% for $x$ axis and 94% for $z$ axis using a Si/SiGe DQD [93]. This is only a partial improvement compared to their result for charge qubits. To make further progress, the detuning point for control should be more positive into spin-like region with longer coherent times and thus Rabi oscillation is preferable. The approach for Rabi oscillations is to set the qubit in spin-like region and apply MW bursts to rotate it around $x$ axis and vary the relative phase of successive MW bursts to rotate it around $z$ axis. An example of the Rabi oscillations is shown in Fig. 3(h). In 2015, Kim et al. applied this method to the hybrid qubit and acquired a control fidelity of 93% for $x$ axis and 96% for $z$ axis [95]. This fully improved qubit quality compared to the charge qubit and simpler control method compared to the spin-1/2 qubit attracts a lot of attentions to transplant the hybrid qubit into other systems. However, this qubit design relies on the valley-orbit states in silicon and thus cannot be borrowed directly. To address this problem, Cao et al. in 2016 implemented this qubit in a region with more electrons, (2, 3)-(1, 4) instead of (2, 1) and (1, 2), in a GaAs DQD [96]. The increased number of electrons allows tuning the mixture of charge and spin degrees freely such that the energy levels can be encoded like in a Si/SiGe DQD. Later, Wang et al. extended the hybrid qubit into a TQD. With an extra quantum dot for energy level tuning, they realized a tunable operation frequency from 2 to 15 GHz, allowing a large range for frequency multiplexing [97]. In fact, the valley splitting in Si/SiGe quantum dots are not so controllable and varies from sample to sample. These new types of hybrid qubit are free of valley states and thus are more reproducible and scalable.

## Two qubit gate in semiconductor

In contrast to single qubit gates that all require two-axis control, the two qubit gate can be realized in many different ways. In fact, the CNOT gate is not the only two qubit gate for universal quantum computing. Others also include the square root of swap gate ($\sqrt{SWAP}$) and the controlled phase gate (CZ) [16]. The SWAP gate swaps two qubit state and the $\sqrt{SWAP}$ gate performs half-way of the SWAP gate. The CZ gate acts on two qubits in such a way that a π rotation around $z$ axis is performed on the target qubit only when the control qubit state is $|1\rangle$. In the semiconductor quantum devices, these different two qubit gates can also be divided into three different categories considering the source of interaction: exchange interaction, Coulomb interaction and circuit quantum electrodynamics (cQED). In the following subsections, we will introduce the realization of two qubit gates using different types of interactions and discuss the progresses.

## Exchange interaction

Exchange interaction is a quantum mechanical effect for identical particles. In this context, it refers to the interaction between two spins. The two qubit gate using exchange interaction have been proposed for spin-1/2 qubits [20], singlet-triplet qubits [62], exchange-only qubits [76] and hybrid qubits [91]. Among them the exchange interaction between spin-1/2 qubits were investigated most thoroughly in experiments and thus we mainly discuss it in the following. The interaction strength $J$ and Zeeman energy difference $\Delta E_z$ are two competing factors in controlling two interacting spins, and their relative magnitude determines the energy levels of the system. Fig. 4(a) depicts the energy level spectrum in four different cases [98]: (I) When both $\Delta E_z$ and $J$ equal zero, the qubit eigenstates are directly product states and all single spin flip transitions are energetically degenerate. (II) If only $\Delta E_z$ is nonzero, two spins can be addressed at different transition frequencies and single spin qubit control can be achieved. (III) If $\Delta E_z$ and $J$ are nonzero and $J$ is much bigger than $\Delta E_z$, the two qubit eigenstates are no longer effectively product states but singlet and triplets. It is just like the case of singlet-triplet qubit, and a $\sqrt{SWAP}$ gate can be implemented with a $\pi/2$ rotation around $z$ axis. (IV) If $\Delta E_z$ and $J(\varepsilon_s)$ are nonzero and $J$ is much smaller than $\Delta E_z$, the qubit eigenstates can still be viewed as product states with small corrections due to spin-charge hybridization. In this regime, each qubit transition frequency is no more independent of the state of the other and thus permits CZ or CNOT operations.

For the $\sqrt{SWAP}$ gate, it was first demonstrated by Petta et al. using GaAs quantum dots in 2005, reporting an operation on input state $|\uparrow\downarrow\rangle$ or $|\downarrow\uparrow\rangle$ with time of 180 ps [44]. However, limited by the measurement method, they could not perform the $\sqrt{SWAP}$ gate for other input states like $|\uparrow\uparrow\rangle$ or $|\downarrow\downarrow\rangle$. In 2011, Nowack et al. first demonstrated independent single-shot readout of two electron spins using energy selective readout, and upon this result they measured the full truth table for a SWAP gate with four different input states [99]. In the same year, Brunner et al. combined the $SWAP^n$ gate ($n$ means multiples of the operation time of a SWAP gate) with single qubit rotations and demonstrated a two qubit entanglement [100].

For the CZ gate in semiconductor, it was first theoretically discussed by Meunier et al. in 2011 [101]. The energy levels as functions of detuning $\varepsilon_s$ is shown in Fig. 4(b), a vanishing detuning lowers the antiparallel-spin states with $J(\varepsilon_s)/2$ and thus allows a phase shift of $J(\varepsilon_s)t_{wait}/2$ when applying a detuning pulse for a fixed time $t_{wait}$, resulting in a unitary transformation in the basis of $|\uparrow\uparrow\rangle$, $|\uparrow\downarrow\rangle$, $|\downarrow\uparrow\rangle$ and $|\downarrow\downarrow\rangle$:

$$U_{Cphase} = \begin{pmatrix} 1 & 0 & 0 & 0 \\ 0 & e^{iJ(\varepsilon_s)t_{wait}/2} & 0 & 0 \\ 0 & 0 & e^{iJ(\varepsilon_s)t_{wait}/2} & 0 \\ 0 & 0 & 0 & 1 \end{pmatrix} \quad (10)$$

When $t_{wait}$ equals $\pi/J(\varepsilon_s)$, this gate control correspond to a CZ gate only with additional single qubit $z$ rotations. The CZ gate was first demonstrated in a silicon MOS DQD by Veldhorst et al. in 2015 [102]. By combining it with two $\pi/2$ rotations, they implemented a CNOT gate and observed the corresponding anti-correlations. In 2018, Watson et al. used dynamical decoupling pulses to improve the performance of CZ gates and performed the Deutsch-Josza algorithm and the Grover search algorithm with a natural Si/SiGe DQD, suggesting the first implementation of a programmable two qubit quantum processor [103]. The Bell state tomography, which is a characterization of the two qubit gate performance, indicated prepared state fidelities of 85-89%. Considering the state preparation and measurement (SPAM) errors brought about by the Bell state tomography, they then used character randomized benchmarking to study the CZ gate control fidelity and obtained a value of ~91% [104].

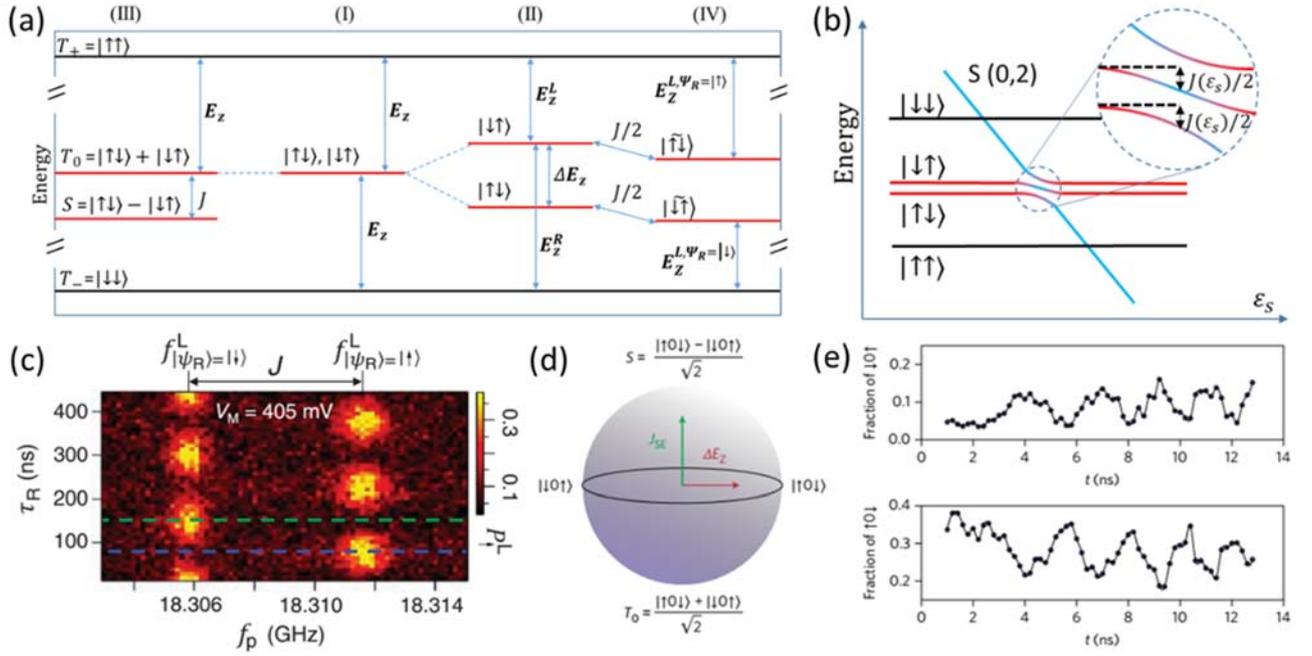

**Figure 4.** Two qubit gates based on exchange interaction. (a) Eigenenergies of two spins in a DQD in the presence of a magnetic field gradient $\Delta E_z$ and relevant transitions between them for four distinct realistic parameter regimes: (I) both $\Delta E_z$ and $J$ equal zero, (II) only $\Delta E_z$ is nonzero, (III) both $\Delta E_z$ and $J$ are nonzero and $J$ is much bigger than $\Delta E_z$, (IV) both $\Delta E_z$ and $J$ are nonzero and $J$ is much smaller than $\Delta E_z$. (b) Energy levels of two spin states as a function of detuning $\varepsilon_s$ in condition (IV). The energy shift $J(\varepsilon_s)/2$ of the antiparallel-spin states is denoted in the enlarged dotted circle. (c) The probability of spin-up states for the left qubit $P_\uparrow^L$ as a function of the MW burst time $\tau_R$ and MW frequency $f_p$. The MW bursts are applied on the right qubit. Two resonance frequencies of the left qubit are split by $J$ (Adapted from ref. [17]). (d) Bloch sphere representation of the singlet-triplet subspace in the superexchange regime with control axes $J_{SE}$ and $\Delta E_z$ (Adapted from ref. [107]). (e) Observation of superexchange-driven spin oscillations (Adapted from ref. [107]).

For the CNOT gate, it can be realized by directly driving the qubits via MW bursts for a time when $J$ is nonzero [17, 34]. As Fig. 4(a) suggests, MW bursts with a frequency resonant with the transition of $|\downarrow\uparrow\rangle$ to $|\uparrow\uparrow\rangle$ and off-resonant with other transitions can cause the left qubit to rotate only when the right qubit state is $|\uparrow\rangle$. As a result, the rotation of the left qubit is controlled by the right qubit state, and it corresponds to a CNOT gate when the controlled rotation angle equals π, as illustrated in Fig. 1(b). Actually, the CNOT gate here has to be calibrated to eliminate the conditional phase caused by exchange interaction, and usually we call it conditional rotation (CROT) gate. In experiments, $J$ can be controlled by manipulating the detuning $\varepsilon_s$ or the inter-dot tunneling $t$. In Watson et al.'s experiment, they controlled $\varepsilon_s$ to implement a CROT gate for measuring a qubit state via another qubit [103]. In 2018, Zajac et al. realized a direct CNOT gate via controlling inter-dot tunneling $t$, reporting a Bell state fidelity of 78% [17]. The device is shown in Fig. 1(c), and they used the middle gate M to directly control the inter-dot tunneling and thereby the exchange interaction. When the interaction is turned on, the resonance frequency of the left qubit is dependent on the right qubit state. As illustrated in Fig. 4(c), the response of the left qubit to MW bursts oscillates between two frequencies as the right qubit is under Rabi oscillation, and the two state dependent resonance frequencies are separated by $J$. On top of that, the CROT gate can also be implemented with a constant $J$. With this new approach, in 2018, Huang et al. set up a new record with fidelity up to 98% via two qubit randomized benchmarking based on a purified silicon MOS DQD [34].

In addition, except two nearest qubits, a two qubit gate on the strength of exchange interaction can also be applied to a qubit with the next nearest neighbor. This was investigated by a number of groups. In 2013, both Braakman et al. and Busl et al. found a direct tunnel coupling of two outer quantum dots of a TQD, suggesting a superexchange interaction may exist between the two electron spins in the outer quantum dots [105, 106]. With the empty central quantum dot acting as a mediator, in 2016, Baart et al. first demonstrated superexchange interaction driven oscillations of two distant spins in the outer quantum dots [107]. The Bloch sphere representation of the $S$-$T_0$ states of the two outer spins and corresponding $S$-$T_0$ oscillations are shown in Fig. 4(d) and Fig. 4(e), respectively. Similar to singlet-triplet qubits, the $z$ axis of the Bloch sphere is controlled by the superexchange interaction $J_{SE}$ while the $x$ axis is controlled by the Zeeman energy difference $\Delta E_z$. Moreover, a multi-electron quantum dot can also serve as a mediator for long range exchange interaction and in 2018 Malinowski et al. demonstrated exchange oscillations of two distant spins using this method [108]. Another approach to realize coupling with a next nearest neighbor is to shuttle a local entangled state to a farther quantum dot. In 2018, Nakajima et al. found that this approach can be assisted by dephasing noise and observed corresponding coherent evolutions [109]. Apart from these, the exchange interaction can be applied to different types of qubits. In 2018, a CZ gate was implemented for a spin-1/2 qubit and a singlet-triplet qubit via a tunable inter-qubit exchange coupling by Noiri et al. [110] in a GaAs TQD.

As for donor spin-1/2 qubits, the exchange coupling is not that easy to harness for its rigid requirement of atomic-scale precision of two donors. A recent progress in 2018 was the observation of anti-correlated spin states between two donors in silicon separated by 16±1 nm [111]. The small tunnel coupling in their device prohibited measurement of coherent oscillations, indicating a much smaller separation is needed for further research. To resolve this challenge, a series of two qubit gate strategies with slightly relaxed requirements on donor placement were proposed, including taking advantage of hyperfine interaction [112] and magnetic dipole-dipole coupling [35], or introducing an intermediate coupler like probe spin [36] and quantum dots [37]. Moreover, the recent proposed flip-flop qubit promises a two qubit gate that can be implemented at separations of hundreds of nanometers with the second-order electric dipole-dipole interaction [113]. Nonetheless, there is still a long way ahead for these proposals to come true.

## Coulomb interaction

Coulomb interaction is the electrostatic coupling between two or more electrons. The two qubit gate based on Coulomb interaction has been proposed for singlet-triplet qubits [114], exchange-only qubits [115], hybrid qubits [92] and charge qubits [116]. Until now, both the experiments on singlet-triplet qubits and charge qubits have been demonstrated.

For singlet-triplet qubits, the two qubit gate was first experimentally investigated by Weperen et al. in 2011 using two electrostatically coupled DQDs [117]. They found that the precession frequency of the singlet-triplet qubit can be controlled by the charge arrangement of an electrostatically coupled DQD. Then, in 2012, Shulman et al. utilized the different charge occupations of $S$ and $T_0$ to control the precession frequency of another qubit [118]. A typical device and a schematic of the charge configurations of $S$ and $T_0$ are shown in Fig. 5(a) and (b), respectively. The Hamiltonian can thus be given by

$$\boldsymbol{H}_{2ST} = (J_1(\sigma_z \otimes I) + J_2(I \otimes \sigma_z))/2 + (\Delta E_{z,1}(\sigma_z \otimes I) + \Delta E_{z,2}(I \otimes \sigma_z))/2 + J_{12}(\sigma_z - I) \otimes (\sigma_z - I)/4 \quad (11)$$

Here $I$ is the identity operator, $J_i$ and $\Delta E_{z,i}$ are the exchange splitting and Zeeman energy difference, with $i = 1, 2$ referring to the corresponding qubit, and $J_{12}$ is the two-qubit coupling strength. When rotating both qubits around $z$ axis simultaneously, the state dependent charge occupation will mediate a coupling strength $J_{12} \propto J_1 J_2$, and an entangled state may be produced with a certain operation time. With dynamical decoupling pulses, they implemented an entangling gate in this way and measured the generated state, yielding a Bell state fidelity of 72%. Here, the entangling gate is equivalent to a CZ gate up to single-qubit rotations. Using a dominating magnetic gradient to suppress charge noise, Nichol et al. reported an entangling gate fidelity of 90% in 2017 [71].

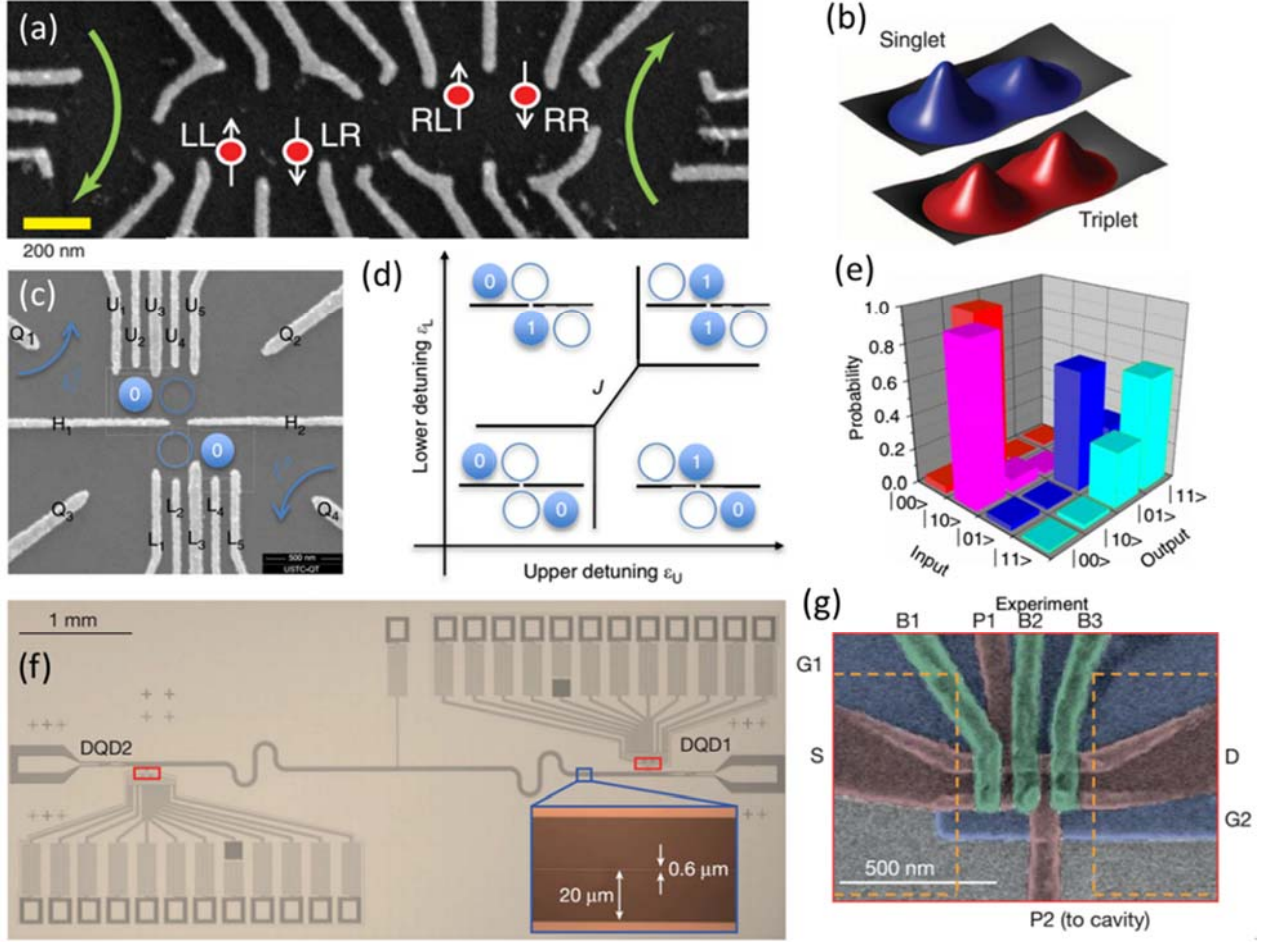

**Figure 5.** Two qubit gates based on Coulomb interaction and circuit quantum electrodynamics (cQED) with a DQD. (a) SEM image of a device for entangling two singlet-triplet qubits. The approximate locations of the electrons in the two qubits are denoted by red circles with arrows. The current paths for the SETs are denoted by green arrows (Adapted from ref. [118]). (b) Schematic of the charge configuration for singlet (blue) and triplet (red) (Adapted from ref. [118]). (c) SEM image of a device for coupling two charge qubits. The solid blue circles denote the charge configuration for the corresponding state (Adapted from ref. [120]), and two current paths for QPCs are denoted by blue arrows. (d) Diagram showing Coulomb interaction induced $J$ as a function of detuning for the upper DQD $\varepsilon_U$ and lower DQD $\varepsilon_L$ (Adapted from ref. [120]). (e) Probabilities for the output states of a CROT operation acquired experimentally by preparing qubits in different input states $|00\rangle, |10\rangle, |01\rangle$ and $|11\rangle$ (Adapted from ref. [120]). (f) Optical image of the superconducting resonator coupling two DQDs. The inset shows an optical image of the center pin and vacuum gap (Adapted from ref. [126]). (g) False color SEM image of a DQD with gate P2 coupled to the resonator. The micro-magnets for spin-photon coupling are indicated by orange dashed lines (Adapted from ref. [126]).

For charge qubits, it was first investigated by Petersson et al. [119] and Shinkai et al. [116] in 2009. The level structure of the interacting two qubit system was probed and correlated oscillations were observed. In 2015, on the basis of these results, Li et al. demonstrated a CROT gate of two strongly coupled charge qubits [120]. The two qubit Hamiltonian can be written as,

$$\boldsymbol{H}_{2C} = (\varepsilon_1(\sigma_z \otimes I) + \varepsilon_2(I \otimes \sigma_z))/2 + t_1(\sigma_x \otimes I) + t_2(I \otimes \sigma_x)$$
$$+ J_{12}(\sigma_z - I) \otimes (\sigma_z - I)/4 \quad (12)$$

Here $\varepsilon_i$ and $t_i$ are the detuning and inter-dot tunneling, with $i = 1, 2$ referring to the qubit, and $J_{12}$ again is the two-qubit coupling strength. A typical device [120] is depicted in Fig. 5(c), showing two electrostatically coupled GaAs DQDs. Charge qubits are formed in each DQD and can be controlled independently by detuning. Fig. 5(d) shows the interaction between two charge qubits when controlling the detuning. The zero detuning point for the upper qubit, i.e. the anti-crossing point for qubit to change from $|0\rangle$ to $|1\rangle$, will shift to a higher point by an amount of $J_{12}$ when the lower qubit state is changed from $|0\rangle$ to $|1\rangle$. Here we denote the lower point as $\varepsilon_U^{\Psi_L = |0\rangle}$ and the higher point $\varepsilon_U^{\Psi_L = |1\rangle}$. A CNOT gate can thus be applied by pulsing the upper qubit to $\varepsilon_U^{\Psi_L = |0\rangle}$ so that the upper qubit will be rotated only if the lower qubit state is $|0\rangle$. In this way, they measured the truth table (see Fig. 5(e)) of a CROT gate and extracted a control fidelity ~68%. In 2016, Ward et al. also demonstrated controlled rotations for two charge qubits in two coupled Si/SiGe DQDs. To take a step further, in 2018, Li et al. demonstrated three-qubit controlled rotations using three coupled GaAs DQDs [121], which is a first attempt to go beyond the two-qubit limit in semiconductor devices and suggests the semiconductor qubits are amenable to large-scale manufacture.

## Coupling to the resonator

In addition to proximity coupling like exchange interaction and Coulomb interaction, semiconductor qubits can also be coupled distantly through cQED [122, 123]. In 2012, both Frey et al. and Petersson et al. demonstrated a DQD dipole coupled to an on-chip distributed superconducting resonator [124, 125]. An experiment setup and the corresponding DQD with a surface electrode connected to the resonator are shown in Fig. 5 (f) and (g), respectively [126]. For the cQED, a Hamiltonian of the Jaynes-Cummings type can be given as

$$\boldsymbol{H}_{DC} = \omega_c(n + 1/2) + \omega_q \sigma_z/2 + g_c \sin\theta \, (a^\dagger \sigma^- + a\sigma^+) \quad (13)$$

Here the first term refers to the photon of the resonator with photon frequency $\omega_c$ and photon number operator $n$, the second term refers to the DQD with transition frequency $\omega_q = \sqrt{\varepsilon_c^2 + 4t^2}$, the third term refers to the photon-DQD coupling with coupling rate $g_c$, the creation and annihilation operators of photon $a^\dagger$ and $a$, the Pauli operators $\sigma^-$ and $\sigma^+$, and $\sin\theta = 2t/\sqrt{\varepsilon_c^2 + 4t^2}$. A change of detuning $\varepsilon_c$ and tunneling $t$ of the DQD will alter the Hamiltonian and cause a modification of photons in the resonator, resulting in the phase and amplitude change of the transmitted or reflected signal, which can be observed in experiments. It is noteworthy that the SQD can also be coupled to a resonator [127] and due to space limitation we will not go into detail for its mechanism.

Using photons in the resonator as a mediator, distant coupling of two DQDs are thus expected. After 2012, several groups reported nonlocal coupling between distant quantum dots using a resonator, including Delbecq et al. [127] and Deng et al [128]. However, the relatively strong atomic (qubit) dephasing rate ϒ, which is inversely proportional to the dephasing time, and photon loss rate κ impeded their step for two qubit gate control. To overcome this challenge, a strong coupling regime in which $g_c$ exceeds $\kappa$ and $\Upsilon$ is on demand. Since 2016, the strong coupling of a resonator to a DQD in the Si/SiGe heterostructure [129], the GaAs/AlGaAs heterostructure [130] and the carbon nanotube [131] have been reported successively. Upon these results, in 2018, Nicolí et al. [132] demonstrated tunable photon-mediated coupling of two charge qubits and measured the two-qubit coupling strength. In the meanwhile, Scarlino et al. [133] also demonstrated coherent coupling between a semiconductor charge qubit and a superconductor qubit, and even observed controlled oscillations. On top of these results, strong spin-photon coupling were realized in 2018 by several groups [126, 134, 135]. Further exploration on coherent coupling and even a two qubit gate of spin-1/2 qubits [136], singlet-triplet qubits [137] and resonant exchange qubits [138] are thus expected.

## Challenges and opportunities

The developments and recent advances of high-fidelity single- and two-qubit controlin semiconductor have been introduced above, indicating a strand of a scalable fault-tolerant quantum computing fabric. However, there are still some challenges to resolve before a next leap. Here, we discuss these challenges and introduce some research progress.

## Readout of qubits

As discussed above, the readout method of most types of semiconductor qubits depends on a proximate charge sensor. Once the charge state of the quantum dot or a donor changes, the resistance of the charge sensor will change accordingly. Therefore, the readout speed and fidelity is directly related to the bandwidth and signal-to-noise-ratio (SNR) of the charge sensor. To include both the bandwidth and SNR, in the following we use the charge sensitivity to characterize the performance of a readout method:

$$\text{Charge sensitivity} = 1/((\text{SNR}) \cdot \sqrt{\text{Bandwidth}}) \ (e/\sqrt{\text{Hz}}) \tag{14}$$

A typical state-of-the-art charge sensor with a transconductance amplifier at room temperature (RT) can achieve a charge sensitivity down to 820 $\mu e/\sqrt{\text{Hz}}$ for a 30 kHz bandwidth [139]. To improve its performance, several approaches have been investigated [140].

The first approach is to couple an impedance matching radio frequency (rf) resonant circuit to the integrated charge sensor, usually a SET or a QPC, to form a rf-SET [141] or a rf-QPC [142, 143]. Its operating principle is to detect the modulation of the reflected or transmitted rf signal by resistance change of the charge sensor. The impedance matching network lowers the high resistance of the charge sensor, usually > 50k$\Omega$, towards the characteristic impedance of a transmission line ~50 $\Omega$. Thus the RC time constant of the circuit is reduced and thereby the working bandwidth is improved. The first demonstration of using a rf-SET to detect charge states in semiconductor was in 2003, when Lu et al. fabricated an aluminum rf-SET to detect real-time electron tunneling in a GaAs quantum dot [141]. In 2007, Reilly et al. and Cassidy et al. reported the characterization of rf-QPCs fabricated using the GaAs/AlGaAs heterostructure [142, 143]. Both the rf-SET and the rf-QPC can offer a charge sensitivity lower than 200 $\mu e/\sqrt{\text{Hz}}$ with a bandwidth over 1 MHz. For the applications in qubit readout, in 2009, Barthel et al. used a rf-QPC to detect a singlet-triplet qubit and reported a single-shot measurement with fidelity over 90% for a bandwidth ~143 kHz [144]. The rf-QPC they used is depicted in Fig. 6(a), with an Ohmic contact (box) coupled to an impedance match work formed by an inductor and a parasitic capacitance of the bond pads and wires. In 2010, by using a GaAs quantum dot based rf-SET, they even improved the measurement bandwidth to 10 MHz for the readout of singlet-triplet qubit [145]. Beyond that, this technique also applies to other types of qubits such as charge qubits [146], spin-1/2 qubits [18] and exchange-only qubits [77].

The second approach is to use cryo-amplifiers. For conventional measurement method, the readout bandwidth is also limited by the transconductance amplifier at room temperature (RT). When the readout bandwidth is increased, the RT amplifier gain will decrease and so does the SNR. In fact, the SNR can still be increased if the amplification is introduced at a lower temperature before the dominant noise comes in. Inspired by this idea, several attempts have been made to fabricate a suitable cryo-amplifier located much closer to the device, including employing a high electron mobility transistor (HEMT) [147] and a SiGe herterojunction bipolar transistor (HBT) [148]. In 2016, Tracy et al. demonstrated the single-shot readout of a P donor electron spin-1/2 qubit with 96% visibility of Rabi oscillations by using a cryogenic two stage HEMT circuit adjacent to the qubit device [149], as shown in Fig. 6(b). For a bandwidth of 100 kHz, they achieved a SNR of 9 and a charge sensitivity of 300 $\mu e/\sqrt{Hz}$, nearly twice as good as the values of the state-of-the-art charge sensor with a RT amplifier.

Different from the approaches mentioned above, the third approach gets rid of a charge sensor by coupling a surface electrode of a DQD directly to a resonator. The mechanism has been introduced in the section for two qubit gates. For experiments, in 2010, Petersson et al. first demonstrated a lumped-element resonator circuit coupled to the reservoir of a DQD in GaAs and used it to probe charge and spin states [150]. Then in 2013, Colless et al. coupled the lumped-element resonator to a gate electrode of a DQD and named it rf gate sensor[151]. A rf gate sensor is shown in Fig. 6(c), with a lumped-element resonator constituted by an inductor $L$~210 nH and a parasitic capacitance $C_p$~ 0.2 pF. In 2015, Gonzalez-Zalba et al. demonstrated a record sensitivity of 37 $\mu e/\sqrt{Hz}$ for a bandwidth ~ 1kHz with a gate sensor [152] for silicon SOI quantum dots. This value was further improved to 1.3$\mu e/\sqrt{Hz}$ in 2018 [153] for a bandwidth of ~ 10 Hz. The single-shot readout of singlet-triplet qubits using rf-gate sensors were realized by Pakkiam et al., Urdampilleta et al. and West et al. in 2018 with donors in silicon, a silicon MOS DQD and a silicon SOI DQD [154-156], respectively. Among them, the best reported readout fidelity is 99.7% for a 1 kHz bandwidth.

Alternatively, the lumped-element resonator can be replaced by an on-chip distributed superconducting resonator, like the experiments for two qubit gates. Early in 2012, Petersson et al. first demonstrated readout of a spin-1/2 qubit using a superconducting resonator based on spin blockade [125]. And in 2015, Stehlik et al. added a Josephson parametric amplifier (JPA) at the output of the resonator to amplify the signal, as shown in Fig. 6(d), resulting in a charge sensitivity of 80 $\mu e/\sqrt{Hz}$ for a bandwidth of 2.6 MHz [157]. Later, Mi et al. replaced the JPA for a traveling wave parametric amplifier (TWPA) and demonstrated strong coupling of the DQD to a resonator [129]. With the help of a slanting magnetic field generated by a micro-magnet, in 2018, they further demonstrated strong spin-photon coupling and performed dispersive readout of a spin-1/2 qubit [126]. For charge qubits, in 2017, Scarlino et al. [158] demonstrated dispersive readout of a charge qubit and measured an intrinsic dephasing time $T_2$ up to (43.1±4.3) ns. Furthermore, coupling to resonators not only requires no charge sensor, but also allows frequency multiplexing. Since 2014, the proposals of multiplexing readout of spin and charge qubits have been demonstrated for larger scale applications [159, 160].

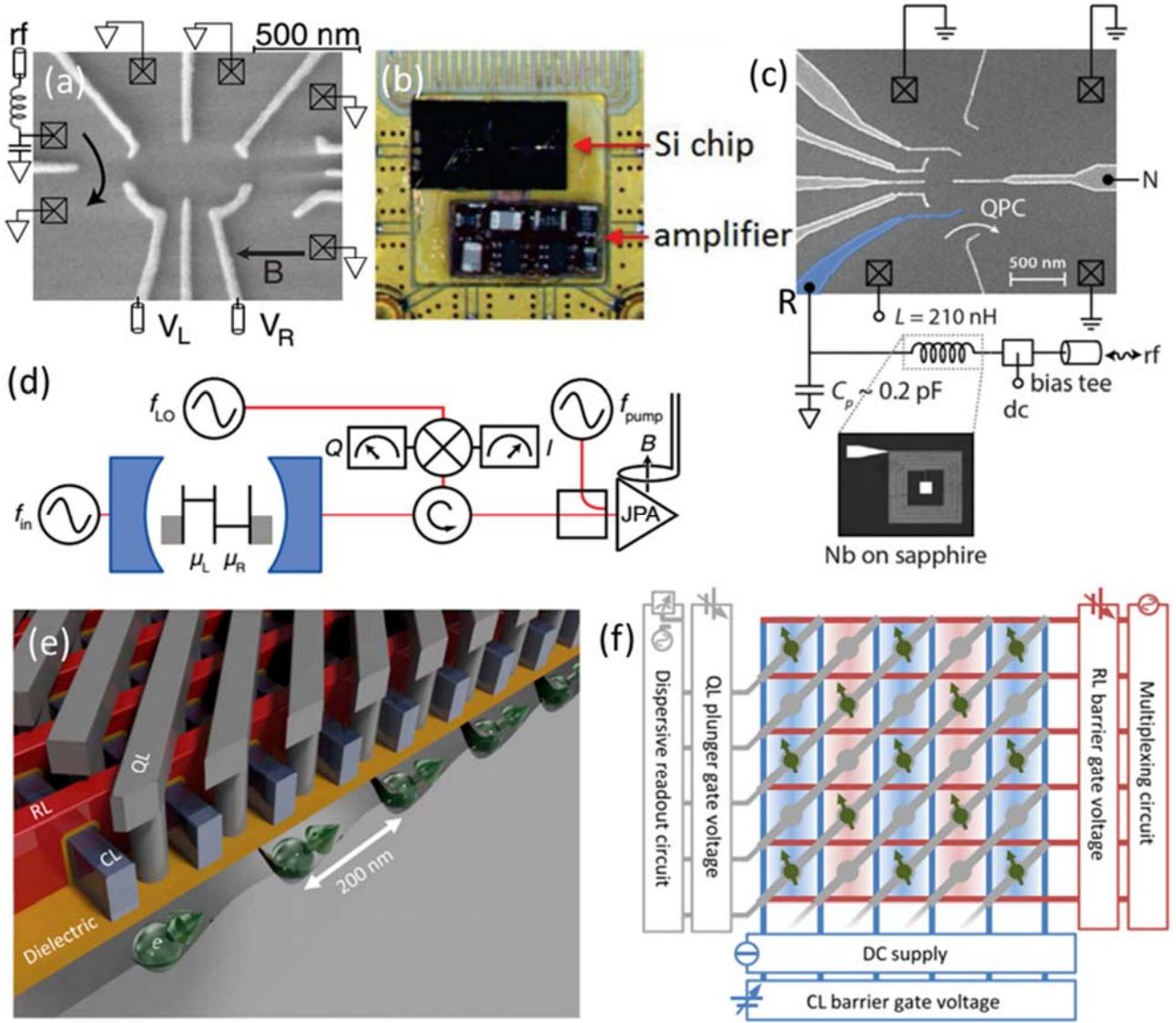

**Figure 6.** Approaches to improve readout quality and a crossbar network for large-scale integration. (a) SEM image of a device using a rf-QPC, indicating Ohmic contacts (boxes), fast gate lines, impedance matching circuit, grounded contacts, and the external magnetic field direction (Adapted from ref. [144]). (b) Picture of a silicon device with an adjacent cryogenic amplifier circuit mounted on printed circuit board (Adapted from ref. [149]). (c) False colored SEM image of a device using a rf gate sensor. One electrode (blue) is coupled directly via a bondwire to an off-chip $Nb/Al_2O_3$ superconducting lumped-element resonator (Adapted from ref. [151]). (d) Diagram of the device using a distributed superconducting resonator with the input field $f_{in}$. The output field is sent to a JPA through a circulator and then demodulated into the I and Q quadratures with a local oscillator tone $f_{LO}$. A directional coupler is used to couple the pump field at frequency $f_{pump}$ (Adapted from ref. [157]). (e) and (f) are three dimensional model and schematic representation of the crossbar network for a 2D quantum dot array. CLs (blue), RLs (red), and QLs (gray) connect the qubit grid to outside electronics for control and readout [39].

## Material developments

The semiconductor qubit has long been argued for excellent scalability considering the success of semiconductor technology for classical computers. However, the largest number of qubits controlled in the same device till now is still no more than four [161]. It can be partially explained by the limited experimental conditions of college in contrast to the industry, but more importantly, the properties of the substrates plays a key role in fabrication. Traditional modulation doped GaAs/AlGaAs heterostructure [21] is an excellent substrate for quantum dots owing to its relatively high mobility and steady quality. However, the nuclear noise in GaAs hinders the research on high fidelity spin qubits. To eliminate spin noise, researchers have to move to another substrate: silicon [17, 103] or even purified silicon with less Si$^{29}$ [18, 53, 102]. Nevertheless, there are many challenges to implement qubits in silicon, such as complicated valley degeneracy and elaborate control of small electronic wave function [22].

The valley degeneracy originates from the six degenerate minima of the conduction band of bulk silicon and these subbands are termed as valleys. For donors in silicon, the additional degeneracy of the valley state is not a concern since the strong confinement potential from the dopant atom can lift it easily. But for silicon quantum dots, although four in-plane valleys are raised far away from the two out-of-plane valleys, the small splitting of two low lying valleys still complicates the qubit control [22]. For spin-1/2 qubits, it has been observed that spin-valley mixing can cause a sizable decrease of spin lifetime [162], the g-factor is renormalized by SOC with valley states [163], the probability of occupying the valley state will deteriorate spin initialization, and inter-valley scattering may limit the dephasing time [52]. For singlet-triplet qubits and exchange-only qubits, the lower valley may lift spin blockade and reduce the readout fidelity [68]. And for hybrid qubits in silicon quantum dots, the energy scale caused by the valley splitting should be controlled in a reasonable range. Therefore, a reproducible and controllable valley splitting in silicon is on demand. Till now, several researches have been performed on valley splitting in silicon [162, 164-168]. As a whole, the valley splitting in quantum dots based on silicon MOS and SOI are in the range of 300-800 μeV and 610-880 μeV, respectively, and can be easily controlled by electric field [162, 167]. A recent research even suggested that the single-electron valley splitting in silicon MOS quantum dots is both tunable and predictable, thus silicon MOS is a promising platform for qubit control [165]. In contrary, the valley splitting in Si/SiGe quantum dots has been found to vary considerably from sample to sample with a range of only 35-270 μeV [166]. The origins may be attributed to the interfacial steps and atomic scale disorder of the Si well and such disorder mainly arises from the substrate and relaxed buffer [168]. To further control the valley splitting in a reproducible way, other researches on how to control such disorder are needed.

For the small electronic wave-function, a much smaller lithographical size of the quantum dot is required for fine control. The traditional depletion mode gate design used in GaAs quantum dots, as shown in Fig. 3(a), (d) and (g), is like a big conference hall and the formed quantum dots are like conferees and it thus puts an obstacle for precise control. In the contrary, a new overlapping gate design that accumulate electrons only under the small electrodes with a diameter on the similar scale of a quantum dot are more preferred. This new design was first introduced by Angus et al. in 2007, who used two layers of aluminum electrodes with local insulator in between to define a quantum dot in silicon MOS, resulting in a lithographical diameter as small as 50 nm [169]. It's worth noting that the lithographical size of the quantum dot refers to the surface area defined by the electrodes and does not equal to the actual size of the quantum dot formed by electric field. The lithographical diameter in silicon MOS was later reduced to 30-40 nm for controlling single electrons [170, 171]. For quantum dots in the Si/SiGe heterostructure, similar electrode designs were demonstrated by Zajac et al. [172] and Borselli et al. [173] in 2015. One device fabricated by Zajac et al. is shown in Fig. 1 (c) with three layers of electrodes, which are confinement gates (blue), plunger gates (red) and barrier gates (green), to form a DQD in the upper channel and a SET in the lower channel. In contrast to silicon MOS devices, the electrode size to form a quantum dot in Si/SiGe heterostructure can be a little larger, usually of ~80 nm. This less stringent requirement for lithography allows a great improvement of reproducibility and scalability of quantum dots. In 2016, Zajac et al. presented a device containing a linear array of 9 quantum dots with reproducible properties and 3 quantum dots as SETs [139]. Using a device with the same electrode structure, in 2018, Mills et al. demonstrated shuttling a single charge across this linear array [174]. In contrast, the largest number of working quantum dots in a silicon MOS device is still no more than four [53, 171], implying the difficulty set by lithography. However, other designs including a single layer electrode layout using poly-Si in silicon MOS [175] and silicon quantum dots fabricated using SOI technology were also developed [55], implying alternatives for reproducibility and scalability.

Beyond silicon, germanium, which can host aforementioned hole spin qubits, is also an promising material for high fidelity qubit control. Like silicon, it can be isotopically purified to improve coherence times. Also, it is free of valley degeneracy and thus supports the reproducibility of well-defined qubits. Recent reports on the quantum dots in Ge hut wires and the Ge/GeSi heterostructure have shown some initial results based on this material [56, 60, 176]. To implement high fidelity single- and two-qubit gates in such material and verify its homogeneity and reproducibility, further in-depth researches are still needed.

## Scalable design

Now that high fidelity control and readout of single- and two- qubit gates in semiconductor have been demonstrated, the next challenge lies in how to scale it to tens and hundreds of qubits. Corresponding constraints and problems were investigated thoroughly since 2015, including the geometry and operation time constraints [177, 178], engineering configuration for quantum-classical interface [179-182], and even the quantifying of system extensibility [183]. In the light of these discussions, several proposals for scaling up were proposed, varying from the crossbar network [38, 39] for spin-1/2 qubits in silicon MOS quantum dots, the two dimensional lattice of donor qubits in silicon [35, 36], to the hybrid architecture like donor-dot structure [37] and flip-flop qubit structure[113].

In these proposals, the key issue is the wiring strategy of readout lines and control lines for single- and two- qubit gates as well as the balance between feasibility and high quality performance. Here, we take the crossbar network of silicon spin-1/2 qubits for example to illustrate these considerations. In Li et al.'s work [39], as Fig. 6 (e) and (f) shows, three successive layers that play the roles of column barrier lines (CL), row barrier lines (RL), and diagonal plunger lines (PL) form a two dimensional array. Successively tuning CLs, RLs and PLs, electrons can be loaded from reservoirs at the array boundary into the qubit array for single electron occupation. Moreover, CLs also carry direct currents to generate magnetic field gradient $\Delta E_z$ for adjacent columns, while QLs are connected to a dispersive readout circuit to play the role of gate sensors. To perform single qubit rotations, global superconducting striplines above the chip are used to provide in-plane oscillating magnetic fields. A $\sqrt{SWAP}$ gate can be performed by two spins in the adjacent rows of the same column without $\Delta E_z$, while the spin-charge conversion relies on the two spins in the adjacent columns of the same row with $\Delta E_z$, which can constitute a spin-blockade regime with QLs to probe the tunneling event. Qubits can be moved freely along the rows and columns of the grid to perform a two qubit gate or readout with the help of spin shuttling. Also, the spin shuttling can be used to add a controllable phase for a single qubit thus a rotation around $z$ axis is achieved without extra control. However, obviously, the local control of one location may cause unwanted side effects in another location owing to the shared control property of these lines [184]. Another previous design [38] that connects every quantum dot in the grid directly via floating gates and vertical transistors can alleviate this problem. However, as a trade-off, the mediating floating gates and vertical transistors requires more extensive manufacturing developments. More than that, there is still a gap between the proposed architectures and the reproducible quantum dots in current experiments. For example, the proposed dot tuning and charge sensing in a two dimensional grid has not been demonstrated simultaneously in experiments. For future advances, more experiments are needed to fill the gap.

# Conclusion and outlook

In summary, we have introduced single- and two-qubit control of different types of semiconductor qubits and discussed their developments and recent ground breaking progress. It should be noted that the developments of semiconductor qubits are very fast and there are also other new types of semiconductor qubits that are in proposal or under developments. We cannot include them all limited by the writer's knowledge or due to space limitation, but we still hope this review can provide a clear picture of the state-of-the-art works. For quantum computation, we also discuss the challenges and new opportunities in this field, including new readout methods, material developments and scalable designs. These issues are considered most by researches now and corresponding research progress may determine the future directions to some extent. Since the superconducting circuits and ion traps have been sought after by industry for implementing a quantum computer, semiconductor quantum computation is also getting more and more attention by companies and governments in many countries. In fact, several groups have shown their ambition to further optimize semiconductor qubit quality for large-scale integration and scale it up to ten or more qubits in the next ten years, such as Vandersypen group and Veldhorst group at TU Delft in Netherlands with the help of Intel, Simmons group at Centre for Quantum Computation and Communication Technology (CQC2T) in Australia, the French Alternative Energies and Atomic Energy Commission (CEA) in France and etc.

In 2018, professor John Preskill pointed out that noisy intermediate-scale quantum (NISQ) technology will be available in the near future [185]. This terminology refers to a quantum computer with 50-100 qubits that can surpass the capabilities of today's classical computers [186] and also a low circuit depth that is limited by insufficient control fidelity. Although they will not change the world immediately, the near-term applications for NISQ devices will still be consequential and could bring new opportunities for research and business. From this point of view, in addition to further improving the qubit quality and striving for fault tolerant quantum computing in the coming years, the researches on new applications for semiconductor devices are also important. Therefore, we anticipate the semiconductor quantum devices may develop fast and have a great impact on our lives from now on.

# Acknowledgement:

This work was supported by the National Key Research and Development Program of China (Grant No.2016YFA0301700), the National Natural Science Foundation of China (Grants No. 61674132, 11674300 and 11625419), the Strategic Priority Research Program of the CAS (Grant XDB24030601), the Anhui initiative in Quantum information Technologies (Grants No. AHY080000) and this work was partially carried out at the USTC Center for Micro and Nanoscale Research and Fabrication.